\documentclass[12pt]{elsarticle}
\usepackage{natbib}
\usepackage{amsmath}
\usepackage{xcolor}

\journal{Chaos, Solitons and Fractals}

\begin{document}

\begin{frontmatter}

\author[URJC]{Gaspar Alfaro}
\ead{gaspar.alfaro@urjc.es}
\author[URJC,Kaunas]{Miguel A.F. Sanjuán\corref{cor1}}
\ead{miguel.sanjuan@urjc.es}
\cortext[cor1]{Corresponding author}

\affiliation[URJC]{organization={Nonlinear Dynamics, Chaos and Complex Systems Group, Departamento de  Física, Universidad Rey Juan Carlos},
	addressline={Tulip\'an s/n}, 
	city={Móstoles},
	postcode={28933}, 
	state={Madrid},
	country={Spain}}

\affiliation[Kaunas]{organization={Department of Applied Informatics, Kaunas University of Technology},
	addressline={Studentu 50-415}, 
	city={Kaunas},
	postcode={LT-51368}, 
	country={Lithuania}}      

\title{Time-dependent effects hinder cooperation on the public goods game}

\begin{abstract}
	
The public goods game is a model of a society investing some assets and regaining a profit, although can also model biological populations. In the classic public goods game only two strategies compete: either cooperate or defect; a third strategy is often implemented to asses punishment, which is a mechanism to promote cooperation. The conditions of the game can be of a dynamical nature, therefore we study time-dependent effects such an as oscillation in the enhancement factor, which accounts for productivity changes over time. Furthermore, we continue to study time dependencies on the game with a delay on the punishment time. We conclude that both the oscillations on the productivity and the punishment delay concur in the detriment of cooperation.

\end{abstract}



\end{frontmatter}

\section{Introduction} \label{intro}

Interactions between individuals and the formation of groups have indeed complex dynamics. The interactions of a few may have a very different behavior than when more individuals are accounted for, then emergent properties may arise, and definitely its study gets in the area of complex systems dynamics. Studies on these matters \cite{SocialPhy} include social dilemmas and games, populations grow over time and certain aspects of their behaviour evolve. Taking into account all these ingredients, the Evolutionary Game Theory (EGT) has been developed. EGT involves evolutionary dynamics and aims to understand biological populations, social and economical dynamics through evolving populations. Individuals get payoffs according to their strategy as well as the strategy of others, and change their strategy dynamically. One of the main focus of study is to understand how an altruistic behaviour between individuals appears. Cooperators pay some cost to benefit another individual or the whole, that is why many studies offer ways to promote cooperation. Cooperation has been studied thoroughly on games such as \textit{rock-paper-scissors} \cite{RPSCooperation}, the \textit{prisoner's dilemma} \cite{Prisionero} or the \textit{public goods game} \cite{PublicGoods}.

In the \textit{public goods game} (PGG), individuals choose between two strategies: either cooperate or defect. Furthermore, individuals form groups, and as a matter of fact each individual can be part of several groups. On the other hand, cooperators, $C$, contribute an amount of $1$ to the group, whereas defectors, $D$, contribute $0$. Once all contributions are made, the sum in each group is multiplied by a factor $r$ and shared equally among all group members. Under these premises, defectors take advantage of the cooperators and have the greatest payoff. Cooperation only occurs under $r$ values close to the group size~\cite{r/G}. This means that for lower values of $r$ only defectors will be present after some iterations, impoverishing the population with a global payoff of $0$. This is known as the ``tragedy of the commons" and is present in many systems. 

On the classic game, the parameters that define the dynamics of the game stay constant. There has been studies where group members change in time \cite{EdgeRule}, where individuals can choose to disassociate from others if they defect. But not much work has been done on the dynamics of the game when the parameters vary with time. One parameter that is very likely to change is $r$, which could be seen as a measure of the productivity of an activity, where some goods are invested and grow. Knowing that productivity in the real world changes over time, we have studied the dynamics of the PGG under time oscillations of the enhancement factor $r$. For that purpose, we have chosen a sinusoidal oscillation as a simple and illustrative example.

As we have commented earlier, cooperation occurs when the parameter $r$ is large. The  \textit{public goods game} favors defection, but cooperation makes the global payoff higher. As in any society cooperation is risky, but if cooperators join they prosper. Promoting cooperation between individuals will make a society flourish. To that end, mechanisms such as migration \cite{Migration}, reputation \cite{Reputation}, rewarding the cooperators \cite{Reward} or punishing the defectors has been studied throughout different games or dilemmas. Punishing can be done either by social exclusion \cite{SocialExclusion} or by a monetary fine \cite{Punish2}. In this article we focus on the latter by introducing a third strategy, represented by the punishers, $P$. These contribute to the public goods an amount of $1$ and pay an additional amount to punish the defectors with a fine. Even though punishers have a lower payoff than cooperators after paying the cost to punish, they prevail over cooperators, and increase the cooperation rate. This increase in cooperation rate occurs because, in presence of punishers, defection is less profitable, even though it is the wining strategy, and so defectors do not grow as much; so punishers that surround defectors have a chance to reproduce. It would be natural to think that cooperators should grow more than punishers since cooperators have larger payoffs. But a certain number of punishers are needed to keep the defectors from taking over. As seen in \cite{SocialExclusion}, the frequency of punishers has to be larger than a punisher threshold to stop invasion from defectors.

In the cited articles, the punishment fine is charged at the instant of defection, but actions do not have immediate response times in real life. Delay has been accounted for in numerous studies, and it sometimes promotes new dynamics on the system \cite{Delay}. Here we have studied the changes in the dynamics when there is a delay $\tau$ on the moment that the defectors receive its punishment, as studied on ã \cite{Late} on live individuals.

The main observation that we have made is that instability, in the form of great amplitude of oscillation of the enhancement factor, and delay, hinders cooperation. We have also noted that rapid oscillations, like those of noise, do not alter significantly the result.

The structure of this manuscript is as follows. On Section~\ref{model}, we explain the model of strategy evolution and the methods used to simulate the game. On Section~\ref{punish}, we show how the payoffs of each individual are calculated. We develop the punishment on two different ways: pool and peer punishment, and discuss the differences. On Section~\ref{oscillating}, we introduce time oscillations of parameter $r$ and discuss the results. On Section~\ref{delay}, we introduce a delay on the system and discuss the results. And finally, on Section~\ref{Conclusions}, we present the conclusions.

\section{Model of the simulation}
\label{model}

The \textit{public goods game} can be studied with live individuals \cite{Late} or simulated in either spatial games \cite{Punish2,ValorK} or in networks as graphs \cite{EdgeRule}. It can also be studied by deterministic dynamical equations theoretically \cite{SocialExclusion} with averaged payoffs and well-mixed populations.

We have studied the \textit{public goods game} as a spatial game and used a square grid to make PGG simulations. In each cell, there is an individual that plays with its immediate neighbors. That way each individual plays $5$ games on different groups of $G=5$ individuals in each step of the simulation. For each group $g=1,... G$ they play in, they obtain a payoff $\Pi^g$. The accumulated payoff of each individual is the sum of the payoff they got on every game they played that round, $\Pi=\sum_g^G \Pi^g$. On each side of the square grid there are $L$ cells which, unless stated otherwise, we have set to $300$. Therefore, the number of individuals in the simulations is $N=L^2=N_C+N_D+N_P$. 

In all evolutionary games, a evolution mechanism reflecting the payoff of each individual is necessary. Individuals with greatest payoff, i.e.,  more adapted, will reproduce more likely. Here, the evolution mechanism is the imitation rule, and the experiment proceeds as a Monte Carlo simulation. Therefore, in each step of the simulation an individual $x$ is chosen at random to change strategies with one of its neighbors $y$ once they have played their games. The probability to change the strategy depends on the accumulated payoff $\Pi_{x,y}=\sum_g^G \Pi_{x,y}^g$ of both $x$ and $y$ individuals as indicated below.

\begin{equation}
W(s_x \rightarrow s_y)=\frac{1}{1+\exp[(\Pi_{x}-\Pi_{y})/K]},
\end{equation}
$K$ serves as a regulator of the stochastic noise that appears when adopting a strategy. In real life situations, not everyone makes the best possible action. As $K \to 0$ the individuals always change the strategy if their payoff is smaller than the neighbor's. As $K \to \infty$ the probability of change is $1/2$, regardless of the payoff. On accordance with \cite{ValorK}, we set $K=0.5$ as a fully representative value.

Since the individual that changes his strategy is chosen at random at each iteration, after $N$ iterations, every individual has had the opportunity to change once on average. $N$ iterations is our Monte Carlo Step (MCS). A MCS is the evolutionary meaningful time unit that accounts as one generation.

The initial conditions, i.e., the starting strategies of each individual, may vary the final outcome. All the results in this paper correspond to individuals choosing their strategy randomly at the beginning. 

\section{The punishment methods: pool and peer punishment}
\label{punish}

Charging a monetary fine to defectors as a way of punishment is a very effective way of promoting cooperation and also used very widely on real societies. But the act of punishment is not free, that is, the punishers have to pay this cost. One option is to form a public reserve to which they add some amount of money every time they play the PGG, i.e., pool punishment, or they can choose to pay the cost of punishment every time they detect a defector, i.e., peer punishment. On pool punishment, the defectors are charged with the fine only once when they have one or more punishers in its group. On peer punishment each punisher penalizes with the same fine to each defector. This way the defectors are more severely punished. We set the fine as $\beta=0.125$ and the cost of punishment as $\gamma=0.0125$

For pool punishment, the payoff of each strategy is given by

\begin{equation}\small
\begin{split}
&\Pi_C^g=\frac{r}{G}(N_C^g+N_P^g)-1 \\
&\Pi_P^g=\frac{r}{G}(N_C^g+N_P^g)-1-\gamma \\
&\Pi_D^g = \left\{ \begin{array}{ll}
\frac{r}{G}(N_C^g+N_P^g) & \mbox{if $N_P^g=0$} \\
\frac{r}{G}(N_C^g+N_P^g)-\beta & \mbox{if $N_P^g\neq0$}.
\end{array}
\right\}
\end{split}
\end{equation}

On the other hand, for peer punishment, the payoff of each strategy is given by

\begin{equation}
\begin{split}
&\Pi_C^g=\frac{r}{G}(N_C^g+N_P^g)-1 \\
&\Pi_P^g=\frac{r}{G}(N_C^g+N_P^g)-1-\gamma N_D^g \\
&\Pi_D^g=\frac{r}{G}(N_C^g+N_P^g)-\beta N_P^g.
\end{split}    
\end{equation}

\begin{figure}
	\centering
	\includegraphics[width=1\linewidth]{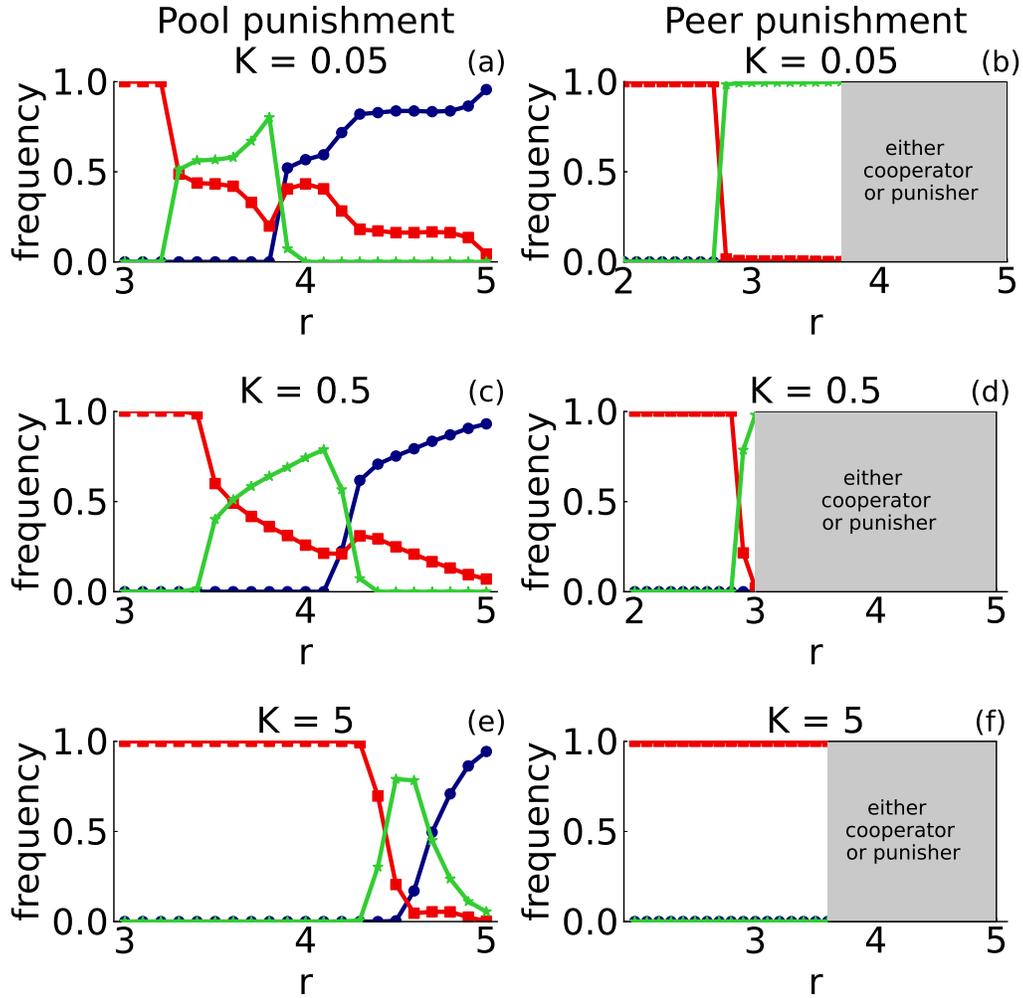}
	\caption{Frequency of each strategy after $5000$ MCS of a PGG simulation as a function of $r$ for different values of the stochastic noise regulator $K$. Cooperators in blue, defectors in red and punishers in green. For greater noise values, cooperation is less profitable. (b) (d) (f) For peer punishment, at some $r$ values defectors become extinct, so  punishers have the same payoff as cooperators. With sufficiently large relaxation times, either punishers or cooperators will go extinct as of neutral drift. (f) For $K=5$ the phase shift between whole domination from defectors and its extinction is very sudden (the step between consecutive $r$ values is $0.1$). The lines are used to guide the eye, and its width is larger than the corresponding error. We have used the following parameter values: $\beta=0.125$, $\gamma=0.0125$}
	\label{densidad}
\end{figure}

The frequency of each strategy after $5000$ MCS is represented on Fig.~\ref{densidad} as a function of the parameter $r$ for different values of the stochastic noise regulator $K$. It can be observed that for low $r$ values, defecting is the wining strategy, but for greater $r$ values, punishers start to dominate. This is because for greater $r$ values the profit of contributing is better. For even greater $r$ values, the punishers lose dominance in exchange for cooperators since they have slightly better payoffs, defectors increase their frequency momentarily due to the lack of punishers, but decrease it as $r$ grows. Noise makes cooperation less effective as we can see the graphs shifting to the right as we increase $K$.

For peer punishment, at some $r$ values defectors become extinct. Since there is no defectors at this point, punishers have the same payoff than cooperators. After enough relaxation time, one of them will go extinct due to a neutral drift. The probability that either strategy wins depends on the frequency of their strategy when defectors gets extinct and probably on the spatial distribution and system size. Comparing this figure with Figs.~\ref{densidad}(a),(c) and (e), we can conclude that peer punishment is more effective at enhancing cooperation if we maintain the same $\beta$ and $\gamma$ values. This is a expected result since for peer punishment defectors are being fined several times and so punishment is more severe.

\section{Varying the enhancement factor as oscillations on time}
\label{oscillating}

The \textit{public goods game} is a simplification of a real society of individuals that invest some goods to get a profit. The profit gained depends on the productivity of the activity which is invested. This productivity may vary over time. If the investment is on the stock-market, the profit depends on the fluctuations of the stock price. To make the PGG simulations respond to this possible fluctuations on time, we introduce a variation on time of the enhancement factor in form of an oscillation $r=r_0+r_1\sin(\frac{2\pi\omega}{L^2}t+\delta)$, where $\omega$ is the oscillation frequency in units of $MCS^{-1}$, $t$ is the time in units of one iteration, and $\delta$ is a time shift that we have set as $0$. We have used a sinusoidal oscillation as an illustrative example since it is a simple periodic function. The oscillation frequency could range from low frequencies representing business cycles such as expansion and recession, to high frequencies representing rapid oscillations that may be due to noise. 

\begin{figure*}
	\centering
	\includegraphics[width=1\linewidth]{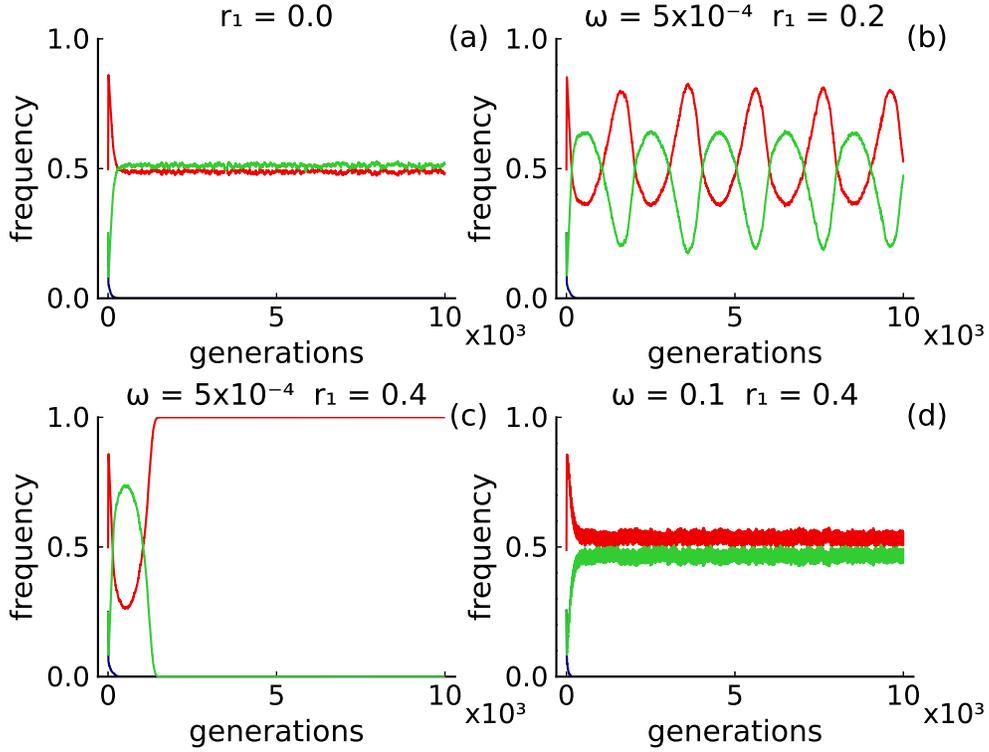}
	\caption{Frequency of each strategy as time passes with an oscillating enhancement factor of the form $r=r_0+r_1\sin(\frac{2\pi \omega}{L^2}t+\delta)$, where  $r_0=3.6$ is the mean value of $r$, $\omega$ is the oscillation frequency in units of MCS$^{-1}$ and $\delta=0$. Results are made in the pool punishment. Cooperators in blue, defectors in red and punishers in green. (a) At the value of $r=3.6$ the punishers and defectors are almost on equal terms rapidly oscillating due to noise. (b) For small oscillation frequencies, the defectors and punishers periodically dominate one another, being the punishers the ones with a greater mean frequency. (c) The amplitude of the oscillation is so big that defectors completely dominate after one cycle. (d) As the oscillation frequency increases, the dynamics is more similar as compared to the case without any $r$ oscillation; while very quick $r$ oscillations, possibly another type of noise, are insignificant to final frequency. We have used the parameter values $\beta=0.125$, $\gamma=0.0125$}
	\label{oscila}
\end{figure*}

The population frequency as time passes is represented on Fig.~\ref{oscila} for different oscillation frequencies and amplitude values with $r_0=3.6$. It can be seen that an increase on the amplitude benefits the defectors, to the extent of absolute domination (Fig.~\ref{oscila}(c)). Thus, we can state that instability hinders cooperation. It can also be seen that for rapid oscillations (Fig.~\ref{oscila}(d)) the dynamics is similar to the one without oscillation, so noise would not affect critically the cooperation rate. This is because the populations do not have time to change significantly on such a short period. All the graphics in the figure are for pool punishment and $K=0.5$, but the conclusions hold for peer punishment and different values of noise.

\begin{figure}
	\centering
	\includegraphics[width=1\linewidth]{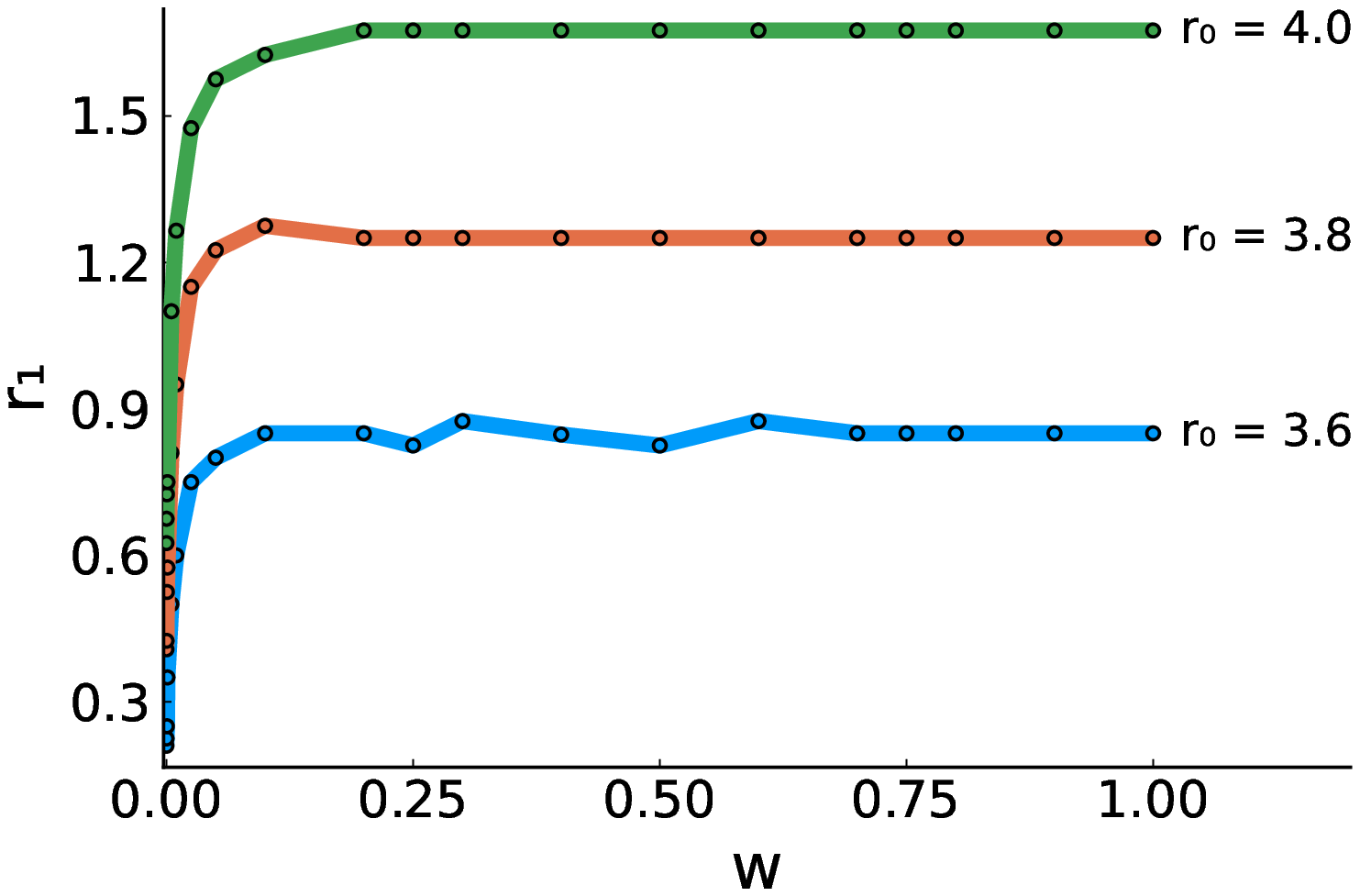}
	\caption{The figure represents the amplitude $r_1$ of the oscillation $r=r_0+r_1\sin(\frac{2\pi\omega}{L^2}t+\delta)$ for different values of $r_0$, that limits the phase of only defectors (greater amplitudes) and defectors plus punishers (lower amplitudes) in the strict pool punishment regime with a punisher threshold $T=0$. The limiting amplitude $r_1$ grows as $r_0$ raises and, at low oscillation frequencies, when $\omega$ raises. The lines are used to guide the eye, and its width is comparable to the corresponding error. We have used the parameters: $\delta=0$, $\beta=0.125$, $\gamma=0.0125$}
	\label{r1_w0}
\end{figure}

As we see on Fig.~\ref{oscila}(c), some combinations of frequency and amplitude of oscillation makes defectors the only lasting strategy. To see for which combinations this occurs, we elaborate Fig.~\ref{r1_w0} by plotting the oscillation amplitude $r_1$ that limits the phase of only defectors and defectors plus punishers versus the oscillation frequency $\omega$ for different values of $r_0$. We can see that the interface is a rising monotonous curve until a limit is reached. At low oscillation frequencies the period when $r$ is large, i.e., when $\sin(2\pi\omega t /L^2 ) \approx 1 $, is large; and there is a lot of time for defectors to grow. That is why the limiting amplitude is very small and increases as $\omega$ grows. It also increases as the mean value of $r$, $r_0$, raises because greater $r$ values benefit cooperators and punishers and so, it is more difficult for defectors to completely dominate. The minimum value of the oscillation frequency plotted is $\omega=0.00001$. For lower values, the ones who dominate the game are the punishers since we have set $\delta=0$ and therefore, at the beginning of the experiment $r$ grows favoring cooperation. 

\section{Having a delay in punishment}
\label{delay}

Information is not instantaneously transmitted and some processes take time to finish. This makes for an stimulus not to have an immediate response. From now on, we suppose that there is a delay of $\tau$ MCS at the time the defectors receive its punishment \cite{Late}. The punishers pay the expense of the punishment the instant they detect someone defecting and the fine is calculated with the conditions of this moment, but actually it is not charged until some time later.

\begin{figure}
	\centering
	\includegraphics[width=1\linewidth]{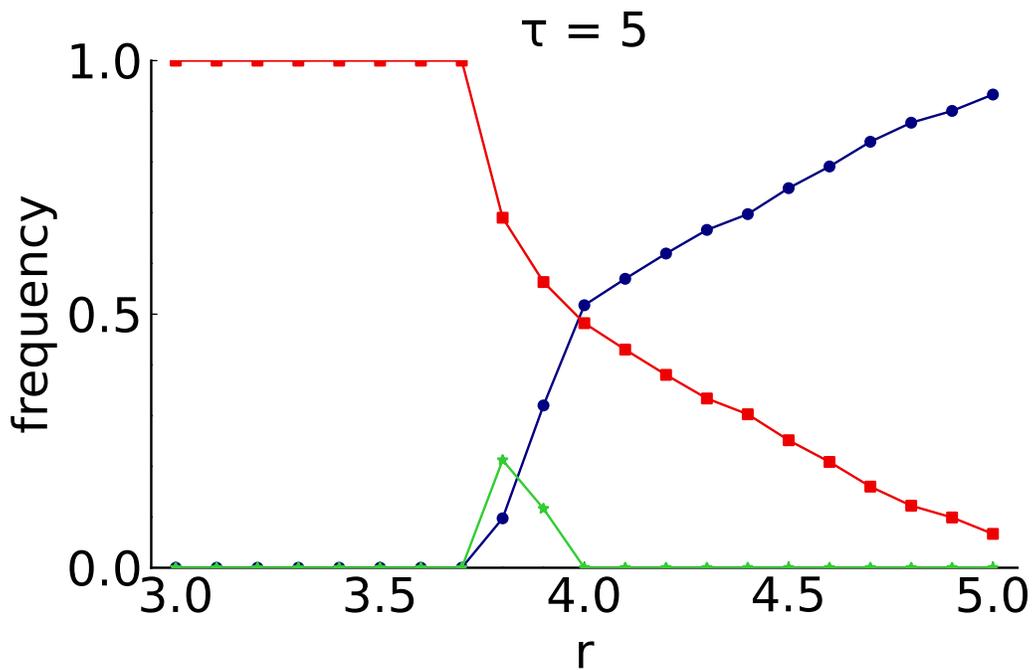}
	
	\caption{
	Frequency of each strategy averaged on the last iterations of a PGG simulation with a delay $\tau=5$ MCS when the defectors are charged with the fine as a function of $r$ in the pool punishment regime. Cooperators in blue, defectors in red and punishers in green. Punishers almost do not appear and defectors are more abundant than on Fig.~\ref{densidad}(c). The lines are used to guide the eye, and its width is larger than the corresponding error. We have used the parameters: $\beta=0.125$, $\gamma=0.0125$, $L=100$. }
	\label{freq_retardo}
\end{figure}

\begin{figure}
	\centering
	\includegraphics[width=0.6\linewidth]{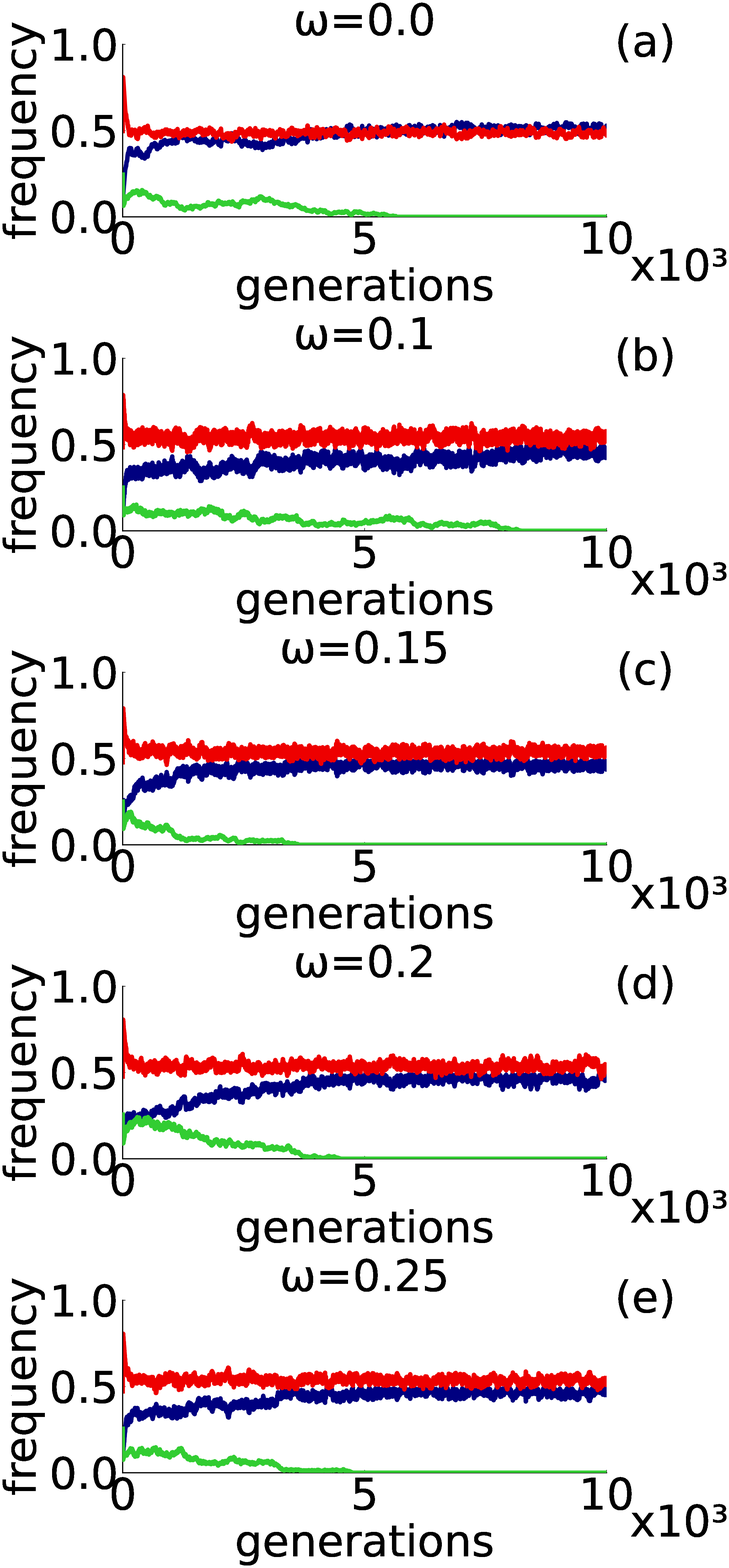}
	\caption{Frequency of each strategy as time passes of a PGG simulation with a delay $\tau=5$ MCS when the defectors are charged with the fine for different oscillation frequencies $\omega$ in the pool punishment regime. (a) No oscillation on the factor $r$. The oscillation of the frequency is due to the delay. With oscillation in the factor $r$, the (b-e) graphics look similar. There is no apparent difference in (d) although $w=0.2=1/\tau$ is the frequency we expected for a resonant behaviour. Cooperators in blue, defectors in red and punishers in green. We have used the parameters: $r_0=4.0$, $r_1=0.5$, $\beta=0.125$, $\gamma=0.0125$}
	\label{tiempo_retardo_oscilacion}
\end{figure}

The frequency of each strategy, averaged on the last iterations, is represented on Fig.~\ref{freq_retardo} as a function of time with a delay $\tau=5$ MCS. This simulation takes longer to calculate, so we have reduced the population ($L=100$). Comparing with Fig.~\ref{densidad}(c), punishers have gone extinct for mostly every value of $r$. Cooperators are on the scene at lower $r$ values but defectors are more abundant than when $\tau=0$, so we can state that the delay hinders cooperation too. This makes sense, since as defectors are charged with the fine later, they have more opportunity to grow. Children have to be punished immediately after the misconduct to know why they are been punished. The same applies here, the larger the punishment delay is, the less effective punishment becomes becomes. This holds for peer punishment and all values of noise.

The evolution of the population frequency as time passes has been represented on Fig.~\ref{tiempo_retardo_oscilacion}. Several simulations with $\tau=5$, $r_0=3.8$, $r_1=0.5$ and different oscillation frequencies are being plotted. The purpose of this figure is to see if there is any type of resonance phenomenon near $\omega=1/\tau=0.2$ (d), but no difference is observed between the graphics of different frequencies. This may be due to the non-periodic behaviour of the population frequency at $\omega=0$. As a matter of fact, there is no periodic oscillation, contrary to what was expected. Moreover, the frequency spectrum does not show a component at a possible delay frequency $0.2$. The same simulations were calculated in the noiseless regime, but a periodic behaviour was not found either. So the reason for this is not due to the noise, but to the intrinsic randomness to the choice of which individual changes their strategy on the Monte Carlo simulation.

\section{Conclusions }\label{Conclusions}

We have studied time-dependent effects on the \textit{public goods game} under the premises of punishment with an additional strategy to the classic game: the punisher. Two ways of punishment were studied, peer and pool punishment. Both punishment strategies proved similar results except for a more effective promotion of cooperation on peer punishment. This is because defectors are being punished multiple times for the same infraction, therefore, the punishment is greater.

To investigate the dynamical nature of conditions of the game, we introduced a time-dependent enhancement factor $r$ and a time delay $\tau$ when the defectors receive its punishment. Both effects bring as a consequence a loss of cooperation. Actually, the population of defectors increases due to the great values of the oscillation amplitude of $r$. Oscillations in productivity hinder cooperation. Furthermore, for high oscillation frequencies, the time series is similar to the case when there are no oscillations. Rapid oscillations, like those given by noise, do not affect greatly the result. On the other hand, when the time delay was present, punishment had little effect and defection was more abundant. Therefore, the punishment delay also hinders cooperation. Finally, we have not observed any resonant behaviour when the two effects are present.

 \section*{Acknowledgments}
This work has been financially supported by the Spanish State Research Agency (AEI) and the European Regional Development Fund (ERDF) under Project No. PID2019-105554GB-I00.

\section*{Conflict of interest}
The authors declare that they have no conflict of interest.

\section*{Credit authorship contribution statement}
\textbf{Gaspar Alfaro:} Investigation, Visualization, Software, Formal analysis, Writing - original draft. \textbf{Miguel A.F. Sanjuán:} Supervision, Conceptualization, Investigation, Formal analysis, Writing - review \& editing, Funding acquisition.

\end{document}